\documentclass{book}
\usepackage{maxent,xspace}

\title {A COMPARISON OF TWO APPROACHES: \\ 
MAXIMUM ENTROPY ON THE MEAN (MEM) AND \\ 
BAYESIAN ESTIMATION (BAYES) FOR INVERSE PROBLEMS}
\author{
{\sc Ali Mohammad-Djafari} \\ 
{\em Laboratoire des Signaux et Syst\`emes (CNRS-ESE-UPS)} \\  
{\em \'Ecole Sup\'erieure d'\'Electricit\'e,} \\ 
{\em Plateau de Moulon, 91192 Gif-sur-Yvette, France.} \\ 
{\em E-mail: djafari@lss.supelec.fr}
}
\def\bdoc{\begin{document}}
\def\edoc{\end{document}}

\def\babs{\begin{abstract}}
\def\eabs{\end{abstract}}

\def\bbib{\begin{thebibliography}{999}}
\def\ebib{\end{thebibliography}}

\def\beq{\begin{equation}}
\def\eeq{\end{equation}}

\def\beqn{\begin{eqnarray}}
\def\eeqn{\end{eqnarray}}

\def\beqnn{\begin{eqnarray*}}
\def\eeqnn{\end{eqnarray*}}

\def\barr{\begin{array}}
\def\earr{\end{array}}

\def\bqu{\begin{quote}}
\def\equ{\end{quote}}

\def\bqun{\begin{quotation}}
\def\equn{\end{quotation}}

\def\bit{\begin{itemize}}
\def\eit{\end{itemize}}

\def\ben{\begin{enumerate}}
\def\een{\end{enumerate}}

\def\bpic{\begin{picture}}
\def\epic{\end{picture}}

\def\bfig{\protect \begin{figure}}
\def\efig{\protect \end{figure}}

\def\bcc{\begin{center}}
\def\ecc{\end{center}}

\def\brr{\begin{flushright}}
\def\err{\end{flushright}}

\def\bll{\begin{flushleft}}
\def\ell{\end{flushleft}}

\def\btab{\begin{tabular}}
\def\etab{\end{tabular}}

\def\ul{\underline}
\def\ol{\overline}

\def\bm#1{\mbox{\boldmath $#1$}}

\def\zerob{{\bm 0}}
\def\oneb{{\bm 1}}

\def\ab{{\bm a}}
\def\bb{{\bm b}}
\def\cb{{\bm c}}
\def\db{{\bm d}}
\def\eb{{\bm e}}
\def\fb{{\bm f}}
\def\gb{{\bm g}}
\def\hb{{\bm h}}
\def\ib{{\bm i}}
\def\jb{{\bm j}}
\def\kb{{\bm k}}
\def\lb{{\bm l}}
\def\mb{{\bm m}}
\def\nb{{\bm n}}
\def\ob{{\bm o}}
\def\pb{{\bm p}}
\def\qb{{\bm q}}
\def\rb{{\bm r}}
\def\sb{{\bm s}}
\def\tb{{\bm t}}
\def\ub{{\bm u}}
\def\vb{{\bm v}}
\def\wb{{\bm w}}
\def\xb{{\bm x}}
\def\yb{{\bm y}}
\def\zb{{\bm z}}

\def\Ab{{\bm A}}
\def\Bb{{\bm B}}
\def\Cb{{\bm C}}
\def\Db{{\bm D}}
\def\Eb{{\bm E}}
\def\Fb{{\bm F}}
\def\Gb{{\bm G}}
\def\Hb{{\bm H}}
\def\Ib{{\bm I}}
\def\Jb{{\bm J}}
\def\Kb{{\bm K}}
\def\Lb{{\bm L}}
\def\Mb{{\bm M}}
\def\Nb{{\bm N}}
\def\Ob{{\bm O}}
\def\Pb{{\bm P}}
\def\Qb{{\bm Q}}
\def\Rb{{\bm R}}
\def\Sb{{\bm S}}
\def\Tb{{\bm T}}
\def\Ub{{\bm U}}
\def\Vb{{\bm V}}
\def\Wb{{\bm W}}
\def\Xb{{\bm X}}
\def\Yb{{\bm Y}}
\def\Zb{{\bm Z}}

\def\alphab{\bm{\alpha}}
\def\betab{\bm{\beta}}
\def\deltab{\bm{\delta}}
\def\epsilonb{\bm{\epsilon}}
\def\gammab{\bm{\gamma}}
\def\omegab{\bm{\omega}}
\def\thetab{\bm{\theta}}
\def\xib{\bm{\xi}}
\def\lambdab{\bm{\lambda}}
\def\phib{\bm{\phi}}
\def\mub{\bm{\mu}}
\def\psib{\bm{\psi}}
\def\chib{\bm{\chi}}
\def\sigmab{\bm{\sigma}}

\def\Deltab{\bm{\Delta}}
\def\Lambdab{\bm{\Lambda}}
\def\Phib{\bm{\Phi}}
\def\Psib{\bm{\Psi}}
\def\Sigmab{\bm{\Sigma}}

\def\Ac{{\cal A}}
\def\Bc{{\cal B}}
\def\Cc{{\cal C}}
\def\Dc{{\cal D}}
\def\Ec{{\cal E}}
\def\Fc{{\cal F}}
\def\Gc{{\cal G}}
\def\Hc{{\cal H}}
\def\Ic{{\cal I}}
\def\Jc{{\cal J}}
\def\Kc{{\cal K}}
\def\Lc{{\cal L}}
\def\Mc{{\cal M}}
\def\Nc{{\cal N}}
\def\Oc{{\cal O}}
\def\Pc{{\cal P}}
\def\Qc{{\cal Q}}
\def\Rc{{\cal R}}
\def\Sc{{\cal S}}
\def\Tc{{\cal T}}
\def\Uc{{\cal U}}
\def\Vc{{\cal V}}
\def\Wc{{\cal W}}
\def\Xc{{\cal X}}
\def\Yc{{\cal Y}}
\def\Zc{{\cal Z}}

\def\wt#1{\widetilde{#1}}
\def\wh#1{\widehat{#1}}
\def\ra{\rightarrow}
\def\la{\leftarrow}
\def\da{\downarrow}
\def\ua{\uparrow}

\def\Ra{\Rightarrow}
\def\La{\Leftarrow}
\def\Da{\Downarrow}
\def\Ua{\Uparrow}

\def\lra{\longrightarrow}
\def\lla{\longleftarrow}
\def\Lra{\Longrightarrow}
\def\Lla{\longleftarrow}

\def\lrarr{\leftrightarrow}
\def\Lrarr{\Leftrightarrow}
\def\udarr{\updownarrow}
\def\Uparr{\Updownarrow}

\def\d#1{\,\mbox{d}#1}
\def\dxdy{\d{x} \d{y}}
\def\dwxdwy{\d{\omega_x} \d{\omega_y}}
\def\dxdydz{\d{x} \d{y} \d{z}}

\def\disp#1{{\displaystyle #1}}
\def\diag#1{\mbox{diag}\left\{#1\right\}}

\def\Prob#1{\mbox{Pr}\left\{#1\right\}}
\def\var#1{\mbox{Var}\left\{#1\right\}}
\def\cov#1{\mbox{Cov}\left\{#1\right\}}
\def\corr#1{\mbox{Corr}\left\{#1\right\}}
\def\trace#1{\mbox{Tr}\left\{#1\right\}}
\def\rang#1{\mbox{rang}\left\{#1\right\}}
\def\det#1{\mbox{d\'et}\left\{#1\right\}}

\def\cosf{\cos \phi}
\def\sinf{\sin \phi}
\def\cost{\cos \theta}
\def\sint{\sin \theta}

\def\sgn{\mbox{sgn}}
\def\sinc{\mbox{sinc}}
\def\rect{\mbox{rect}}
\def\sincf#1{\mbox{sinc}\left(#1\right)}
\def\rectf#1{\mbox{rect}\left(#1\right)}
\def\trif#1{\mbox{tri}\left(#1\right)}
\def\xvec#1#2#3{\left\{#1_#2,\ldots,#1_#3\right\}}

\def\vx{\left[x_1,\ldots, x_n\right]^t}
\def\vz{\left[z_1,\ldots, z_n\right]^t}
\def\vw{\left[\omega_1,\ldots, \omega_n\right]^t}
\def\vxi{\left[\xi_1,\ldots, \xi_n\right]^t}
\def\iii{\int_{-\infty}^{+\infty}}
\def\izi{\int_{0}^{\infty}}
\def\izpi{\int_{0}^{\pi}}
\def\izdpi{\int_{0}^{2\pi}}
\def\intd{\int\kern-.8em\int}
\def\intt{\int\kern-.8em\int\kern-.8em\int}
\def\intg{\int\kern-1.1em\int}
\def\sumd{\mathop{\sum\sum}}

\def\sumi{\sum_{i=1}^{M}}
\def\sumj{\sum_{i=1}^{N}}
\def\sumk{\sum_{k=1}^{K}}
\def\sumn{\sum_{n=1}^{N}}
\def\summ{\sum_{m=1}^{M}}

\def\TA#1{{\cal A}\left\{ {#1} \right\}}
\def\TH#1{{\cal H}\left\{ {#1} \right\}}
\def\TP#1{{\cal P}\left\{ {#1} \right\}}
\def\TR#1{{\cal R}\left\{ {#1} \right\}}
\def\TRa#1{{\cal R}^{\dag}\left\{ {#1} \right\}}
\def\BR#1{{\cal B}\left\{ {#1} \right\}}
\def\TF#1{{\cal F}\left\{ {#1} \right\}}
\def\TFI#1{{\cal F}^{-1}\left\{ {#1} \right\}}
\def\TFn#1#2{{\cal F}_{#1}\left\{ {#2} \right\}}
\def\TFnI#1#2{{\cal F}_{#1}^{-1}\left\{ {#2} \right\}}
\def\Im#1{{\cal I}\mbox{m}\left(#1\right)}
\def\Ker#1{{\cal K}\mbox{er}\left(#1\right)}
\def\Imag#1{\mbox{Im}\left(#1\right)}
\def\Re#1{\mbox{Re}\left(#1\right)}
\def\expf#1{\exp\left[ {#1} \right]}

\def\dfdx#1#2{{\mbox{d} {#1}\over{\mbox{d} {#2}}}}
\def\dfdxd#1#2{{\mbox{d}^2 {#1}\over{\mbox{d} {#2}^2}}}
\def\dfdxt#1#2{{\mbox{d}^3 {#1}\over{\mbox{d} {#2}^3}}}
\def\dfdxn#1#2{{\mbox{d}^n {#1}\over{\mbox{d} {#2}^n}}}
\def\dfdxk#1#2{{\mbox{d}^k {#1}\over{\mbox{d} {#2}^k}}}

\def\dpdx#1#2{{{\partial {#1}\over \partial {#2}}}}
\def\dpdxd#1#2{{{\partial^2 {#1}}\over{\partial {#2}^2}}}
\def\dpdxdy#1#2#3{{{\partial ^2 {#1}}\over{\partial {#2} \partial {#3}}}}

\def\arg{\mbox{arg}}
\def\argmins#1#2{\mbox{arg}\min_{#1}\left\{{#2}\right\}}
\def\argmaxs#1#2{\mbox{arg}\max_{#1}\left\{{#2}\right\}}
\def\argmin#1#2{\mathop{\mbox{arg}\min}_{#1}\left\{{#2}\right\}}
\def\argmax#1#2{\mathop{\mbox{arg}\max}_{#1}\left\{{#2}\right\}}

\def\esp#1{\mbox{E}\left\{ #1 \right\}}
\def\espx#1#2{\mbox{E}_{#1}\left\{ #2 \right\}}

\def\wth#1{\widehat{\widetilde{\phantom{#1}}}\!\!\!\! #1}

\def\lrf{L_{r,\phi}}
\def\fw{\widehat{f}(\omegab)}
\def\fthwxi{\wth{f}(\Omega,\xib)}
\def\fthwfi{\wth{f}(\Omega,\phi)}
\def\ftrfi{\widetilde{f}(r,\phi)}

\def\fwxwy{\widehat{f}(\omega_x, \omega_y)}
\def\wxpwy{(\omega_x \, x + \omega_y \, y)}

\def\wtx{\omegab^t \cdot \xb}
\def\ejwtx{\exp\left[j \omegab^t \cdot \xb\right]}
\def\xitx{\xib^t \cdot \xb}

\def\ftrxi{\widetilde{f}(r,\xib)}
\def\ent{-\int p(x) \, \ln p(x) \d{x}}

\def\mean#1{\left< #1 \right>}
\def\slnhn{\sum_{n=1}^N \lambda_n h_n(\rb)}
\def\slngn{\sum_{n=1}^N \lambda_n g_n(\rb)}
\def\smngn{\sum_{n=1}^N \mu_n g_n(\rb)}
\def\slmhm{\sum_{m=1}^N \lambda_m h_m(\rb)}
\def\vlambda{\bm{\lambda} = [\lambda_1,\ldots,\lambda_n]}

\def\apriori{{\em a priori} }
\def\aposteriori{{\em a posteriori} }

\def\titre#1{\bcc{\Large\bf #1}\ecc}

\def\AMD{Ali Mohammad--Djafari}
\def\LSSa{Laboratoire des Signaux et Syst\`emes 
(CNRS--ESE--UPS) \\ 
\'Ecole Sup\'erieure d'\'Electricit\'e \\ 
Plateau de Moulon, 91192 Gif sur Yvette Cedex, France.}

\def\ME{maximum entropy}
\def\pdf{probability distribution function}
\def\lm{Lagrange multipliers}
\def\fix#1{\phi _#1(x)}
\def\fin{\fix n}
\def\fik{\fix k}
\def\fiz{\fix 0}
\def\sfinz{\sum_{n=0}^N \lambda_n \, \fin}
\def\sfinu{\sum_{n=1}^N \lambda_n \, \fin}
\def\bl{\bm{\lambda}}
\def\bd{\bm{\delta}}
\def\blz{\bl ^0}
\def\gnl{G _n(\bl)}
\def\gnlz{G _n(\blz)}
\def\un{n=1,\dots, N}
\def\nn{n=0,\dots, N}

\def\finn{\fin , \nn}
\def\esfinz{\exp\,\left[ -\sfinz \right] }
\def\esfinu{\exp\,\left[ -\sfinu \right] }
\def\esxm{\exp\,\left[ -\sum_{m=0}^N \lambda_m \, x^m \right] }
\def\efin{\esp \fin }
\def\zl{Z(\bl)}
\def\finxi{\phi _n(x_i)}
\def\snfinxi{\sum_{n=1}^N \lambda_n \finxi}
\def\esnfinxi{\exp \left[ - \snfinxi \right]}
\def\smfinxi{\sum_{i=1}^M \finxi}

\def\ejnw{\exp \left( -j n \omega_0 x \right) }
\def\eejnw{\mbox{E} \left\lbrace \ejnw \right\rbrace}

\def\signed#1{{\unskip\nobreak\hfil\penalty50\hskip2em\mbox{}
\nobreak\hfil\tt#1\parfillskip=0pt \finalhyphendemerits=0 \par}}

\def\uncatcodespecials{\def\do##1{\catcode`##1=12 }\dospecials}
\def\listing#1{\par\begingroup\setupverbatim\input#1 \endgroup}
\newcount\lineno
\def\setupverbatim{\tt \lineno=0
 \obeylines \uncatcodespecials \obeyspaces
 \everypar{\advance\lineno by1 \llap{\sevenrm\the\lineno\ \ }}}
{\obeyspaces\global\let =\ }

\def\defined{\stackrel{\mbox{def}}{=}}
\def\str{\stackrel}

\def\ER{\mbox{I\kern-.25em R}}
\def\EC{\mbox{C\kern-.8em C}}
\def\EZ{\mbox{Z\kern-.55em Z}}
\def\EN{\mbox{N\kern-.8em N}}

\def\singles{
 \abovedisplayskip 12pt plus 3pt minus 9pt
 \belowdisplayskip 12pt plus 3pt minus 9pt
 \abovedisplayshortskip 0pt plus 3pt
 \belowdisplayshortskip 7pt plus 3pt minus 4pt
 \baselineskip 14.4pt
 \lineskip 1pt
 \lineskiplimit 0pt}
\def\oneandhalf{
 \abovedisplayskip 18pt plus 3pt minus 9pt
 \belowdisplayskip 18pt plus 3pt minus 9pt
 \abovedisplayshortskip 0pt plus 3pt
 \belowdisplayshortskip 9.333pt plus 3pt
 \baselineskip 20pt
 \lineskip 2pt
 \lineskiplimit 1pt}

\def\double{
 \abovedisplayskip 24pt plus 3pt minus 9pt
 \belowdisplayskip 24pt plus 3pt minus 9pt
 \abovedisplayshortskip 0pt plus 3pt
 \belowdisplayshortskip 12pt plus 3pt
 \baselineskip 27pt
 \lineskip 3pt
 \lineskiplimit 2pt}

\def\dadb{\d{\alpha}\d{\beta}}

\def\ffbox#1{\fbox{\mbox{\vbox{#1}}}}

\def\rot{\mbox{rot}}
\def\case#1#2#3#4{
    \left\{
           \begin{array}{ll}
            {\displaystyle #1} & {\displaystyle #2} \cr 
            {\displaystyle #3} & {\displaystyle #4}
           \end{array}
    \right. }

\def\beqnarr#1&#2&#3\\#4&#5&#6\eeqnarr{
    \left\{
           \begin{array}{lcl}
            {\displaystyle #1} & #2 & {\displaystyle #3} \\ 
            {\displaystyle #4} & #5 & {\displaystyle #6} 
           \end{array}
    \right. }

\def\pyx{p(\yb|\xb)}
\def\pxy{p(\xb|\yb)}

\newtheorem{definition}{D\'efinition}
\newtheorem{theorem}{Th\'eor\`eme}

\def\ie{{\em i.e.}}
\def\unsdpi{\left(\frac{1}{2\pi}\right)}
\def\unspi{\left(\frac{1}{\pi}\right)}

\begin{document}
\maketitle
\thispagestyle{empty}

\begin{abstract} 
To handle with inverse problems, two probabilistic approaches have 
been proposed: the maximum entropy on the mean (MEM) and the Bayesian 
estimation (BAYES). The main object of this presentation is to 
compare these two approaches which are in fact two different inference 
procedures to define the solution of an inverse problem as the optimizer 
of a compound criterion. 
\end{abstract} 


\section{Introduction}
Inverse problems arises in many areas of science and engineering. 
In fact, rarely, we can measure directly a quantity $x$ and, in 
general, the unobserved interested $x$ is related to the 
measured quantity $y$ via a model. In many area 
this model can be written in the general form $g=\Ac(x)+n$ or 
in the discrete case:
\beq \label{lip}
\yb=\Ab(\xb)+\nb,
\eeq
where $\yb$ stands for the data, $\xb$ for the unknown variables and 
$\nb$ for the errors (modeling and noise). Since Newton and Gauss, 
one tries to define a solution to this problem as the optimizer of 
a criterion, for example the Least Squares (LS):
\beq \label{LS-crit}
\wh{\xb}=\argmins{\xb}{\|\yb-\Ab(\xb)\|^{2}}.
\eeq
But the inverse problems are, in general, ill-posed and 
the LS criterion may not have a unique optimum or this solution 
may be very sensitive to noise. Since Tikhonov \cite{Tikhonov77}, 
the regularization theory became 
the main approach to give a satisfactory solution by defining it as 
the optimizer of a compound criterion:
\beq \label{Reg-crit}
\wh{\xb}=\argmins{\xb}{J(\xb)}\quad \hbox{with}\quad 
J(\xb)=Q(\xb)+\lambda\Omega(\xb)
=\|\yb-\Ab(\xb)\|^{2}+\lambda \|\Db\xb\|^{2} 
\eeq
or in its more general forms \cite{Demoment89}: 
\beq \label{Comb-crit}
\wh{\xb}=\argmins{\xb}{J(\xb)} \quad \hbox{with}\quad 
J(\xb)=Q(\yb-\Ab(\xb))+\lambda\Omega(\xb,\mb).
\eeq
The questions then raised on how to choose the functionals $Q$ and 
$\Omega$ and the regularization parameter $\lambda$ and the default 
solution $\mb$. 

The probabilistic approaches started to give partial answers to this 
request. In particular in the Bayesian estimation approach and the 
maximum a posteriori (MAP) estimate:
\beq
\wh{\xb}=\argmaxs{\xb}{p(\xb|\yb)}
=\argmins{\xb}{-\log p(\xb|\yb)}
=\argmins{\xb}{-\log p(\yb|\xb)-\log p(\xb)},
\eeq
this choice is: 
$Q=-\log p(\yb|\xb)$ and $\lambda \Omega(\xb)=-\log p(\xb)$. 
This approach just pushed a little farther the questions which 
became how to translate our prior knowledge into a 
probability law and how to determine their parameters. 
Even, nowadays, there are many tools for the estimation of the 
hyperparameters \cite{Djafari96a}, 
the main question on how to translate some knowledge about $\xb$ 
into a probability distribution stays without a complete answer. 
The maximum entropy (ME) principle gave partial answers 
\cite{Djafari90a,Djafari92,Djafari96a}. 
See also \cite{Kass94} for an extensive discussed bibliography. 

At the same time, many authors used the ME principle to find unique 
solutions to linear inverse problems by considering $\xb$ as a 
distribution and the data $\yb$ as linear constraints on them. 
Then, assuming that the data constraints are satisfied by a non 
empty set of solutions, a unique solution is chosen by maximizing 
the entropy:
\beq \label{me1}
-\sum_{j} x_{j}\log x_{j} \quad\hbox{or}\quad 
-\sum_{j} x_{j}\log \left[\frac{x_{j}}{m_{j}}-(x_{j}-m_{j})\right],
\eeq
where $\mb$ is default solution. 
See for example \cite{Skilling84,Skilling89} and the cited references. 
However, even if in these methods, thanks to convex analysis and 
Lagrangian techniques, the constrained optimization of 
\ref{me1} can be replaced by an equivalent unconstrained optimization, 
the obtained solutions satisfy the uniqueness condition of 
well-posedness but not the stability one 
\cite{Nashed74,Nashed81,Borwein93a}.  

Recently, some authors \cite{Dacunha90,Gamboa89,Navaza85,Navaza86} 
used the ME principle in a different way 
by considering $\xb$ not as a distribution but as the mean value of 
a random vector $\Xb$ and the data as the constraints on its  
distribution $\d{P(\xb)}$. Then, the ME principle is used to define 
it uniquely and finally the solution $\wh{\xb}$ is defined as the 
expected value of this ME distribution.

Following these authors, some others used, commented and analyzed 
extensively these ideas  
\cite{Shore80,Shore81a,VanCampenhout81,Borwein91a,Csiszar91,Decarreau92},  
\cite{Mukherjee84,OSullivan94,Smith79,Michelot94,Ben-tal88} and  
\cite{Borwein92a,Borwein92b,Borwein93,Decarreau92}. 
However, in all these works, the data $\yb$ were considered as exact 
constraints and the errors on the data were either neglected or 
partially token account of. 
(See however new developments in \cite{Heinrich96}.)

More recently, some authors who were more faced with real 
applications,  
\cite{LeBesnerais93a,Bercher94b,Bercher95,Bercher95a} 
followed the same idea, but by fixing themselves as the objective to 
use these ideas for describing the solution as the optimizer of 
a combined convex criteria such as (\ref{Comb-crit}) and 
more on a constructive way to determine these functionals. 

The objective of this paper is to make a comparison of the Bayesian 
approach which we call hereafter BAYES and the maximum entropy in  
the mean which we refer to as MEM. This comparison is done very 
pragmatically and is based on the understanding of the author who 
does not have pretension to know all the details 
of the both approaches and will be happy to discuss all the following 
discussions with the pro of the approaches.  

\section{Maximum entropy on the mean approach}
\subsection{Basics}
The main references to the basics of this approach are 
\cite{Dacunha90,Gamboa89}. The original idea and first applications 
in crystallography are given in \cite{Navaza85,Navaza86}. 
More details and extensions are given in 
\cite{LeBesnerais93a,Bercher94b,Bercher95,Bercher95a}. 
The mathematical aspects of convex analysis and duality theorems 
are given in 
\cite{Rockafellar70,Rockafellar93,Borwein92a,Borwein92b,Borwein93,Decarreau92}.  

The following resumes the different steps of the approach:

\bit 
\item Consider a set ${\Cc}$, assume that $\xb\in{\Cc}$ and 
define a reference measure $\mu(\xb)$:
\beq
\xb\in{\Cc},\qquad 
\mb=\intg_{\Cc} \xb \d{\mu(\xb)},
\eeq 
where $\mb$ is the mean value of $\xb$ under this reference measure. 

\item Consider $\xb$ as the mean value of a random vector $\Xb$ for which 
you assume a probability distribution $P$: 
\beq
\xb=\espx{P}{\Xb}=\intg_{\Cc} \xb \d{P(\xb)}
\eeq
and the data $\yb$ as exact equality constraints on it:
\beq 
\yb=\Ab \xb=\Ab \espx{P}{\Xb}=\intg_{\Cc} \Ab \xb \d{P(\xb)}. 
\eeq 

\item Determine the distribution $P$ by: 
\beq 
\hbox{maximize}\quad  -\intg_{\Cc} \log  
\frac{\d{P(\xb)}}{\d{\mu(\xb)}} \d{P(\xb)} 
\quad\hbox{s.t.}\quad  \yb=\Ab \xb=\Ab \espx{P}{\Xb}. 
\eeq
The solution is calculated via Lagrangian:
\beqn 
{\Lc}(\xb,\lambdab) 
&=& \intg_{\Cc} \left[ \log \frac{\d{P(\xb)}}{\d{\mu(\xb)}} 
- \sum_{i=1}^M  \lambda_i (y_i-[\Ab \xb]_i) \right] \d{P(\xb)} 
\nonumber \\ 
&=& \intg_{\Cc} \left[ \log \frac{\d{P(\xb)}}{\d{\mu(\xb)}} 
- \lambdab^t (\yb-\Ab \xb) \right] \d{P(\xb)} 
\eeqn 
and is given by:
\beq
\d{P(\xb,\lambdab)}=\expf{\lambdab^t [\Ab \xb]-\log  
Z(\lambdab)}\d{\mu(\xb)}, 
\eeq
where 
\beq
 Z(\lambdab)=\intg_{\Cc} \expf{\lambdab^t [\Ab \xb]} \d{\mu(\xb)}. 
\eeq
The Lagrange parameters are calculated by searching the unique 
solution (if exists) of the following system of non linear equations:
\beq
 \dpdx{\log Z(\lambdab)}{\lambda_i}=y_i,\quad i=1,\cdots,M.
\eeq

\item The solution to the inverse problem is then defined as the expected 
value of this distribution:
\beq
 \wh{\xb}(\lambdab)=\intg \xb \, \d{P(\xb,\lambdab)}.  
\eeq
\eit

These steps are very formal. In fact, it is possible to determine 
$\wh{\xb}(\lambdab)$ in a more direct manner.  Using the following 
notations:
\beq
\sb=\Ab^t\lambdab,\quad  
G^*(\sb)=\log Z(\sb)=\log \intg_{\Cc} \expf{\sb^t \xb} \d{\mu(\xb)},
\eeq
and
\beq
H(\xb)=\max_{\sb}\{\sb^t \xb - G^*(\sb)\}, \quad 
D(\lambdab)=\lambdab^t\yb-G^*(\Ab^t\lambdab)
\eeq
it is shown that:
\beqn 
 \wh{\lambdab}=\argmaxs{\lambdab}{D(\lambdab)} 
 && \hbox{(Dual criterion)} \\
 \wh{\xb}=\argmins{\xb\in{\Cc}}{H(\xb)} \hbox{~~s.t.~~}\yb=\Ab\xb 
 && \hbox{(Primal criterion)}, \\ 
 \wh{\xb}(\sb)=\dfdx{G^*(\sb)}{\sb} 
 && \hbox{(Explicit relation)},
\eeqn 
where:  
\bit
\item Functions $G$ and $H$ depend on the reference measure $\mu(\xb)$;  

\item $D(\lambdab)$ is the dual criterion which is function 
of the data and the function G;

\item $H(\xb)=H(\xb,\mb)$ is the primal criterion which is 
a distance measure between $\xb$ and $\mb$ which means: \\ 
-- $H(\xb,\mb) \ge 0$, \quad and \quad  $H(\xb,\mb) = 0 \quad 
\hbox{iff}\quad \xb=\mb$; \\ 
-- $H(\xb,\mb)$ \qquad is differentiable and convex on $\Cc$;  \\ 
-- $H(\xb,\mb) = \infty$ \quad if $\xb\not\in {\Cc}$.
\eit 
Now, to be able to go a little more in details, let assume that the 
reference measure is separable:

\vspace{-12pt}
\beq
\mu(\xb)=\prod_{j=1}^N \mu_{j}(x_{j})
\eeq
then, we have:
\beq
\d{P(\xb,\lambdab)}=\prod_{j=1}^N \d{P_{j}(x_{j},\lambdab)}
\eeq  
and
\beq
G(\sb)=\sum_{j} g_{j}\left(s_{j}\right), \quad 
H(\xb,\mb)=\sum_{j} h_{j}(x_{j},m_{j}), \quad 
\wh{x}_{j}=g'_{j}(s_{j}). 
\eeq
Replacing $\sb=\Ab^t\lambdab$ we obtain: 
\beq
G(\lambdab)=\sum_{j} g_{j}\left( [\Ab^t \lambdab]_{j}\right), \quad 
H(\xb,\mb)=\sum_{j} h_{j}(x_{j},m_{j}), \quad 
\wh{x}_{j}=g'_{j}\left([\Ab^t\widehat{\lambdab}]_{j}\right), 
\eeq
where $h_{j}$ and $g_{j}$ depend on the reference measure $\mu_{j}$:
\bit
\item $g_{j}$ is the log Laplace transform (Cramer transform) of $\mu_{j}$: 
\[
g(s)=\log \int \expf{s x} \d{\mu(x)}; 
\] 

\item $h_{j}$ is the convex conjugate of $g_{j}$: \quad 
   $\disp{h(x)=\max_{\sb}\{s x -g(s)\}}$.
\eit

Let give some examples: 
\[
\barr{c|c|c|c}
& \mu_{j}(x) & g_{j}(s) & h_{j}(x,m) \\ \cline{1-4}  
\hbox{Gaussian:} 
 & \disp{\expf{- \frac{1}{2}(x-m)^2}}        
 & \disp{\frac{1}{2}(s-m)^2}  
 & \disp{\frac{1}{2}(x-m)^2}  \\
\hbox{Poisson:} 
 & \disp{\frac{m^x}{x!}\expf{-m}} 
 & \disp{\expf{m-s}}
 & \disp{-x\log\frac{x}{m}+m-x} \\ 
\hbox{Gamma: } 
 &  \disp{ x^{\alpha-1} \expf{-\frac{x}{m}}} 
 & \disp{\log (s-m)} 
 & \disp{-\log\frac{x}{m}+\frac{x}{m}-1}          
\earr
\]
When $\mu(\xb)$ is not separable it is very difficult to do the 
calculation more in details, excepted the Gaussian case 
$\mu(\xb)={\Nc}(\mb,\Rb_x)$, 
where we have:
\beq
H(\xb,\mb)=-\frac{1}{2} (\xb-\mb)^t \Rb_x^{-1} (\xb-\mb), \quad  
G(\lambdab)=-\frac{1}{2}\|\lambdab\|^{2}, \quad 
D(\lambdab)=\lambdab^t\yb+\frac{1}{2}\|\lambdab\|^{2}.
\eeq
(See however a new presentation of the method in \cite{Heinrich96} 
trying to extend the method for taking account of the correlations.)

\subsection{Extensions}
How to account for the noise: Two approaches have been developed in 
\cite{LeBesnerais93a,Bercher94b,Bercher95,Bercher95a}:

\bit 
\item Replacing the exact equality constraints $\yb = \Ab \xb$ by the 
following inequalities: 
\beq
|y_i-[\Ab\xb]_i| < \epsilon, \quad \hbox{or}\quad 
\|\yb-\Ab\xb\|^2 < \sigma^2
\eeq
and using the duality relations they showed:
\beq
\left\{\barr{l} 
\disp{
\wh{\xb}=\argmins{\xb}{H(\xb)} 
\hbox{~s.t.~} |y_i-[\Ab\xb]_i| < \epsilon, 
\hbox{~with~} H(\xb)=\sum_{j} h_{j}(x_{j})
}\\ 
\disp{
\wh{\lambdab}=\argmaxs{\lambdab}{\tilde{D}(\lambdab)} 
\hbox{~~with~~} \tilde{D}(\lambdab)=D(\lambdab)+\alpha \|\lambdab\|^2
}
\earr\right.
\eeq
where $\alpha$ depends on $\epsilon$ or on $\sigma^2$. 

\item Replacing  $\yb = \Ab \xb$ by $\yb = \Ab \xb + \nb$ and rewriting it 
as follows: 
\beq
\yb = [\Ab | \Ib] \left[\barr{c} \xb \\ \nb\earr\right] \lra 
\yb=\tilde{\Ab} \, \tilde{\xb}
\eeq
and assuming $\mu(\tilde{\xb})=\mu_x(\xb) \mu_n(\nb)$ they showed: 
\beq
\wh{\xb}=\argmins{\xb\in{\Cc}}{{\Qc}(\yb-\Ab\xb)+\alpha H(\xb)}
\eeq
\beq
\hbox{with}\qquad \qquad 
H(\xb)=\sum_{j=1}^N h_{j}(x_{j}), 
\qquad \hbox{and}\qquad 
{\Qc}(\zb)=\sum_{i=1}^M q_i(z_{j}). 
\eeq 
Here also $h_{j}(x_{j})$ and $q_i(z_i)$ depend on the reference measures  
$\mu_x(\xb)$ and $\mu_n(\xb)$. The determination of $\alpha$ is not 
discussed. 
\eit

\section{Bayesian approach}

\subsection{Basics}
The different steps of this approach are now well-known:
\bit
\item From the observation model and the hypothesis (prior knowledge) 
on the noise derive the likelihood $p(\yb|\xb;\betab)$;

\item From the hypothesis (prior knowledge) on $\xb$ derive the prior law 
$p(\xb|\thetab)$;

\item Apply the Bayes rule to obtain 
$p(\xb|\yb;\betab,\thetab)
=p(\yb|\xb;\betab)\, p(\xb|\thetab) / p(\yb;\betab,\thetab)$;  

\item Define an estimation rule via a cost function 
$c(\xb,\wh{\xb})$ by:
\beq
\wh{\xb}(\yb;\betab,\thetab)=\argmin{\zb}{\intg c(\xb,\zb) 
p(\xb|\yb;\betab,\thetab) \d{\xb}}. 
\eeq
\eit 
Different cost functions give different estimators:
\bit
\item Maximum a posteriori (MAP):
\beq
C(\xb,\wh{\xb})=1-\delta(\xb-\wh{\xb}) \lra 
\wh{\xb}=\argmaxs{\xb}{p(\xb|\yb;\thetab,\betab)}.
\eeq

\item Posterior mean (PM): 
\beq
C(\xb,\wh{\xb})=[\xb-\wh{\xb}]^t \Qb [\xb-\wh{\xb}]^t \lra 
\wh{\xb}=\espx{\xb|\yb}{\Xb}=\intg \xb \, p(\xb|\yb;\thetab,\betab) 
\d{\xb}. 
\eeq

\item Maximum of the Marginal a posteriori (MMAP): 
\beq
C(\xb,\wh{\xb})=\prod_{j} 1- \delta(x_{j}-\wh{x}_{j}) \lra 
\wh{x}_{j}=\argmaxs{x_{j}}{p(x_{j}|\yb;\thetab)}, 
\eeq
where
\beq
p(x_{j}|\yb;\thetab)=\intg p(\xb|\yb;\thetab) 
 \d{x}_1 \cdots \d{x}_{j-1} \cdots \d{x}_{j+1} \cdots \d{x}_n.
\eeq
\eit 
To illustrate this, let consider the case of linear inverse problems
\(\yb=\Ab\xb+\nb\) with the following hypothesis:
\bit 

\item $\nb$ is zero-mean, white and Gaussian: 
$\nb\sim{\Nc}\left(\bm{0},(1/\beta)\Ib\right)$ which leads to:
\beq
p(\yb|\xb,\beta)\propto \expf{-\frac{1}{2}\beta\|\yb-\Ab\xb\|^2}.
\eeq

\item $\xb$ is Gaussian: 
$\xb\sim{\Nc}\left(\xb_0,(1/\theta)\Pb_0\right)$:
\beq
p(\xb|\theta)\propto \expf{-\frac{1}{2}\theta 
[\xb-\xb_0]^t\Pb_0^{-1}[\xb-\xb_0]}. 
\eeq
\eit
Then, using the Bayes rule it is easy to show that 
\beq
\xb|\yb\sim{\Nc}(\wh{\xb},\Pb)
\quad\hbox{with}\quad   
\wh{\xb}=\Pb \Ab^t (\yb-\Ab\xb_0), \quad 
\Pb=\left(\Ab^t\Ab+\lambda\Pb_0^{-1}\right)^{-1}.  
\eeq
The MAP solution is:
\beq
\wh{\xb}=\argmaxs{\xb}{p(\xb|\yb)}=\argmins{\xb}{J(\xb)},
\hbox{~~with~~} 
J(\xb)=Q(\xb)+\lambda\phi(\xb),
\eeq
where
\beq
Q(\xb)=\|\yb-\Ab\xb\|^2, \quad 
\phi(\xb)=\xb^t\Pb_0^{-1}\xb=\|\Db\xb\|, \quad 
\lambda=\frac{\theta}{\beta}
\eeq
Now, relaxing the second hypothesis, i.e; choosing other prior 
laws for $\xb$ we obtain other MAP criteria. Let just note some 
special interesting cases:

\bit
\item A Generalized Gaussian law for $\xb$: 
\beq
p(x_{j})\propto\expf{-(x_{j}-m_{j})^{\alpha}}. 
\eeq
The related MAP criterion becomes:
\beq
J(\xb)=Q(\xb)+\phi(\xb)\quad 
\hbox{~with~}\quad 
\phi(\xb)=\sum_{j} (x_{j}-m_{j})^{\alpha}.
\eeq

\item A Gamma law for $\xb$: 
\beq
x_{j}\sim{\Gc}(\alpha,m_{j})\lra p(x_{j})\propto 
(x_{j}/m_{j})^{-\alpha}\expf{-x_{j}/m_{j}}. 
\eeq
The related MAP criterion becomes:
\beq
J(\xb)=Q(\xb)+\phi(\xb)\quad 
\hbox{~with~}\quad 
\phi(\xb)=\sum_{j} \alpha \log\frac{x_{j}}{m_{j}} + \frac{x_{j}}{m_{j}}.
\eeq

\item A Beta law for $\xb$: 
\beq
x_{j}\sim{\Bc}(\alpha,\beta)\lra p(x_{j})\propto x_{j}^{\alpha} 
(1-x_{j})^{\beta}. 
\eeq
The related MAP criterion becomes:
\beq
J(\xb)=Q(\xb)+\phi(\xb)\quad 
\hbox{~with~}\quad 
\phi(\xb)=\alpha \sum_{j} \log x_{j}+\beta \sum_{j} \log(1-x_{j}). 
\eeq

\item A Poisson law for $\xb$: 
\beq
p(x_{j})\propto \frac{m_{j}^{x_{j}}}{x_{j}!} \expf{-m_{j}}. 
\eeq
The related MAP criterion becomes:
\beq
J(\xb)=Q(\xb)+\phi(\xb)\quad 
\hbox{~with~}\quad 
\phi(\xb)=-\sum_{j} x_{j}\log \frac{x_{j}}{m_{j}}+(x_{j}-m_{j}). 
\eeq

\item Markovian models for $\xb$: 
\beq
J(\xb)=Q(\xb)+\phi(\xb)\quad 
\hbox{~with~}\quad 
\phi(\xb)=\alpha \sum_{j}\sum_{i\in N_{j}} V(x_{j},x_i). 
\eeq
\eit

\subsection{Extensions}
The Bayesian approach can be exactly applied when all the direct 
(prior) probability laws ($p(\yb|\xb,\betab)$ and $p(\xb|\thetab)$) 
are assigned. Even, choosing an appropriate law is done in general by 
hand, another difficulty is to determine their parameters 
$(\betab,\thetab)$. This problem has been addressed by many authors 
and the subject is an active area in statistics. 
See \cite{Hall87,Hebert92,Johnson91,Titterington85},   
\cite{Younes88,Younes89,Bouman94,Fessler93,Liang92} 
and also \cite{Fortier93,Djafari93a,Djafari95b}.     

All these methods can mainly be divided in three main families:

\bit 
\item Generalized MAP: In this approach one tries to estimate 
both the hyperparameters and the unknown variables $\xb$ directly 
from the data by defining:
\beq
(\wh{\xb},\wh{\thetab},\wh{\betab})
=\argmaxs{(\xb,\thetab,\betab)}{p(\xb,\thetab,\betab|\yb)}   
\eeq
where 
\beq
p(\xb,\thetab,\betab|\yb)
\propto p(\yb|\xb,\betab) \, p(\xb|\thetab) \, p(\thetab) \, 
p(\betab)   
\eeq
and where $p(\thetab)$ and $p(\betab)$ are appropriate prior laws. 
Many authors used the non informative prior law for them.  

\item Marginalization: In this approach one tries to estimate 
first the hyperparameters by marginalizing over the unknown variables 
$\xb$: 
\beq
p(\thetab,\betab|\yb)\propto p(\betab) \, p(\thetab) \intg 
p(\yb|\xb,\betab)\, p(\xb|\thetab) \d{\xb}
\eeq
and then, using them in the estimation of the unknown 
variables $\xb$:  
\beq
(\wh{\thetab},\wh{\betab})=\argmaxs{(\thetab,\betab)}{p(\thetab,\betab|\yb)}  
\eeq

\item Nuisance parameters: In this approach the hyperparameters 
are considered as the nuisance parameters, so, marginalized:
\beq
p(\xb|\yb)=\intg p(\yb,\xb,\thetab,\betab)\d{\thetab}\d{\betab}
\eeq
and the unknown variables $\xb$ are estimated by: 
\beq
\wh{\xb} =\argmaxs{\xb}{p(\xb|\yb)}
\eeq
To see some more discussions and different possible implementations of 
these approaches see \cite{Djafari95b}. 
\eit 

\section{Discussed points in each approach}

\subsection{{\sc Mem}:}

\bit
\item Choice of ${\Cc}$ and $\mu(\xb)$: 

\noindent -- 
${\Cc}$ must be a convex set, such as: \quad $\ER^N$, 
\quad $\ER_+^N$, \quad $[a,b]^N$  

\noindent --
Up to now, the whole analysis can be done for separable measures.

\noindent --
The only reference measures $\mu_{j}$ which permits to go 
through all the steps are those for which we have analytical expression 
for the Laplace transform of their logarithms. 

\item Accounting for the noise: \\ 
In the first approach only the support and the energy of the 
noise is used.  
In the second approach we have more choices via the reference 
measures $\mu_n(\nb)$, but determination of $\alpha$ stays adhoc. 
In fact, in general, when the reference measures $\mu_n(\nb)$ and 
$\mu_x(\xb)$ depend on any parameters, this approach lacks any tool 
to determine them. 

\item Effective calculation of the solution: \\ 
No problem, and more, this is probably the main interest of the 
approach which defines the solution, by construction, as the 
optimizer of a convex criterion. 

\item Characterization of the solution: \\ 
A sensitivity analysis has been proposed by \cite{Bercher94b}, but, 
in my opinion, this is not enough to characterize a solution to an 
inverse problem. \\ 
It is not easy to use the notions of variance or covariance of 
the solution, because this approach does not define a posterior 
distribution for the solution.  

\item Possibility of the extension of the approach: \\ 
I have not yet seen any extension of this approach to 
non linear inverse problems, or linear inverse problems in which the 
operator depend on unknown parameters, such as blind deconvolution or 
antenna array processing;\\ 
The fact that we have to choose a convex set $\Cc$ on which 
the solution is defined excludes the inverse problems in which we 
know a priori that the solution is discrete-valued (binary, $n$ary 
images for example). This excludes the use of this approach in image 
segmentation or communication inverse problems (canal equalization, 
blind deconvolution, etc.). 
\eit

\subsection{{\sc Bayes}:}

\bit
\item Choice of $p(\xb|\betab)$: \\ 
$p(\xb|\betab)$ can be chosen separable or not. 
Evidently, separable $p(\xb|\betab)$ (Entropic prior laws) simplifies 
the calculations. Accounting for correlations is easily done via 
Markovian models;  
In both cases (Entropic or Markovian prior laws), 
there are some tools for choosing them either by physical 
considerations, or by scale invariance arguments 
\cite{Djafari91,Djafari93c,Djafari93d,Brette94a,Brette94b}. 

\item Choice of the cost function or equivalently of an estimator MAP, 
PM, MMAP: \\  
This choice is done more on the basis of cost calculation. 
MAP calculation needs, in general, global optimization, but 
does not need any integration. \\ 
MP or MMAP needs multidimensional integration, so in general, 
greater cost. However, there are approximate calculation techniques 
based on Monte Carlo methods and Gibbs sampling. 

\item Effective calculation of different solutions: \\ 
For MAP estimate, when the posterior law is unimodal, we can 
use any gradient descent based method, but if this is not the case, 
there are two categories of methods: Simulated Annealing or 
Deterministic relaxation (GNC). For more discussions on Bayesian 
calculations see \cite{Djafari96d} in this volume. 
\eit

\section{Comparisons and discussions}

The following main items are discussed: 

\bit
\item In MEM, the unknowns $\bm{x}$ are considered as the mean values 
of a random vector $\bm{X}$ for which a prior probability measure 
$\d{\mu(\bm{x})}$ is defined. 

\item In BAYES, the unknowns $\bm{x}$ are considered as a sample of a 
random vector $\bm{X}$ for which a prior probability measure 
$p(\bm{x})$ is defined. 

\item In MEM, a probability distribution $p(\bm{x})$ is defined as the 
minimizer of the Kullback distance $K(p,\mu)$ subject to the data 
constraints, and the solution is defined to be $\hbox{E}_p(\bm{X})$. 
What is interesting here is that this solution can equivalently be 
obtained as the minimizer of a convex criterion $J(\bm{x})$ subject 
to the data constraints, and what is more attractive is that, thanks 
to the convex analysis, this solution can also be obtained as the 
stationary point of a dual criterion which can easily be calculated 
numerically. 

\item In BAYES, the posterior law $p(\bm{x}|\bm{y})$ is calculated 
using the Bayes' rule. In fact, the data $\bm{y}$ are considered as a 
sample of a random vector $\bm{Y}$ for which we can define a 
conditional probability law $p(\bm{y}|\bm{x})$ which, when 
used in conjunction with the prior $p(\bm{x})$ in the Bayes' rule 
will give us the posterior law, from which we can define an estimator. 
One of these estimators is the posterior mean $\hbox{E}_p(\bm{X})$, 
but others can also be defined. This posterior law is used not only 
to define an estimate (a solution), but also to calculate any other 
probabilistic information about that solution. 

\item In MEM, in its original version, the data are considered 
as the exact linear constraints. The uncertainty on the data are not 
considered, and consequently, the uncertainty on the solution is not 
handled. However, some extensions are recently presented to take 
account of the errors on the data and to calculate the sensitivity of 
the solution to these errors. 

\item In BAYES, the errors are naturally considered through 
$p(\bm{y}|\bm{x})$ and the uncertainty of the solution through the 
posterior probability $p(\bm{x}|\bm{y})$. 
Naturally then, we can compare the information content of the data 
and the prior model using their entropies. We can also measure 
the relative information content of the posterior to prior model 
by $K(p(\xb|\yb),p(\xb))$. 

\item In MEM, even in their extended versions, it is not easy to handle 
with the hyperparameters. In BAYES, there are the necessary tools to 
handle them. 

\item In MEM, one can not yet handle with non linear problems. 
This is not the case of the BAYES. 
\eit

As a final conclusion, we have to mention that, 
even if the two approaches are different, they can, 
in some cases result to the same definition of the solution as the 
minimizer of the same criterion, and consequently, to give exactly 
the same numerical solutions to a given inverse problem. 
However, we can give different interpretations to the obtained 
result depending on the approach used to reach it. 
The main objective of this paper was to give a succinct presentation 
of the two approaches for the resolution of the inverse problems. 

It is important to note that the two approaches give different 
views and interpretations which can be used advantageously for any 
application. Also, even the Bayesian approach is now really mature, 
the MEM approach is more recent. So, many of the conclusions I made 
today may be altered in future. In particular, new presentation of 
the method by Heinrich et al \cite{Heinrich96} in this volume will 
probably give new possibilities to the MEM method and will push the 
limits of the method to greatest generality.

{\small 
%
%
%
\def\AA{Astrononmy and Astrophysics}
\def\AAP{Advances in Applied Probability}			
\def\ABE{Annals of Biomedical Engineering}
\def\AT{Annales des T\'el\'ecommunications}
\def\AMS{Annals of Mathematical Statistics}
\def\AISM{Annals of Institute of Statistical Mathematics}
\def\AO{Applied Optics}
\def\AP{The Annals of Probability}
\def\AST{The Annals of Statistics}
\def\BMK{Biometrika}
\def\CPAM{Communications on Pure and Applied Mathematics}
\def\EMK{Econometrica}
\def\CRAS{Compte-rendus de l'acad\'emie des sciences}
\def\CVGIP{Computer Vision and Graphics and Image Processing}
\def\GJRAS{Geophysical Journal of the Royal Astrononomical Society}
\def\GSC{Geoscience}
\def\GPH{Geophysics}
\def\GRETSI#1{Actes du #1$^{\mbox{e}}$ Colloque GRETSI} 
\def\CGIP{Computer Graphics and Image Processing}
\def\ICASSP{Proceedings of IEEE ICASSP}
\def\ICEMBS{Proceedings of IEEE EMBS}
\def\ICIP{Proceedings of the International Conference on Image Processing}
\def\ieeP{Proceedings of the IEE}
\def\ieeeAC{IEEE Transactions on Automatic and Control}
\def\ieeeAES{IEEE Transactions on Aerospace and Electronic Systems}
\def\ieeeAP{IEEE Transactions on Antennas and Propagation}
\def\ieeeASSP{IEEE Transactions on Acoustics Speech and Signal Processing}
\def\ieeeBME{IEEE Transactions on Biomedical Engineering}
\def\ieeeCS{IEEE Transactions on Circuits and Systems}
\def\ieeeCT{IEEE Transactions on Circuit Theory}
\def\ieeeC{IEEE Transactions on Communications}
\def\ieeeGE{IEEE Transactions on Geoscience and Remote Sensing}
\def\ieeeGEE{IEEE Transactions on Geosciences Electronics}
\def\ieeeIP{IEEE Transactions on Image Processing}
\def\ieeeIT{IEEE Transactions on Information Theory}
\def\ieeeMI{IEEE Transactions on Medical Imaging}
\def\ieeeMTT{IEEE Transactions on Microwave Theory and Technology}
\def\ieeeM{IEEE Transactions on Magnetics}
\def\ieeeNS{IEEE Transactions on Nuclear Sciences}
\def\ieeePAMI{IEEE Transactions on Pattern Analysis and Machine Intelligence}
\def\ieeeP{Proceedings of the IEEE}
\def\ieeeRS{IEEE Transactions on Radio Science}
\def\ieeeSMC{IEEE Transactions on Systems, Man and Cybernetics}
\def\ieeeSP{IEEE Transactions on Signal Processing}
\def\ieeeSSC{IEEE Transactions on Systems Science and Cybernetics}
\def\ieeeSU{IEEE Transactions on Sonics and Ultrasonics}
\def\ieeeUFFC{IEEE Transactions on Ultrasonics Ferroelectrics and Frequency Control}
\def\IJC{International Journal of Control}
\def\IJCV{International Journal of Computer Vision}
\def\IJIST{International Journal of Imaging Systems and Technology}
\def\IP{Inverse Problems}
\def\ISR{International Statistical Review}
\def\IUSS{Proceedings of International Ultrasonics Symposium}
\def\JAPH{Journal of Applied Physics}
\def\JAP{Journal of Applied Probability}
\def\JAS{Journal of Applied Statistics}
\def\JASA{Journal of Acoustical Society America}
\def\JASAS{Journal of American Statistical Association}
\def\JBME{Journal of Biomedical Engineering}			
\def\JCAM{Journal of Computational and Applied Mathematics}	
\def\JCAT{Journal of Computer Assisted Tomography}
\def\JEWA{Journal of Electromagnetic Waves and Applications}	
\def\JMO{Journal of Modern Optics}
\def\JNDE{Journal of Nondestructive Evaluation}		        
\def\JMP{Journal of Mathematical Physics}
\def\JOSA{Journal of Optical Society America}
\def\JP{Journal de Physique}
\def\JRSSA{Journal of the Royal Statistical Society A}
\def\JRSSB{Journal of the Royal Statistical Society B}
\def\JRSSC{Journal of the Royal Statistical Society C}
\def\JSPI{Journal of Statistical Planning and Inference}        
\def\JTSA{Journal of Time Series Analysis}                      
\def\JVCIR{Journal of Visual Communication and Image Representation} 
	\def\MMAS{???} 
\def\MNAS{Mathematical Methods in Applied Science}
\def\MNRAS{Monthly Notes of Royal Astronomical Society}
\def\MP{Mathematical Programming}
	\def\NSIP{NSIP}  
\def\OC{Optics Communication}
\def\PRA{Physical Review A}
\def\PRB{Physical Review B}
\def\PRC{Physical Review C}
\def\PRD{Physical Review D}				
\def\PRL{Physical Review Letters}			
\def\RGSP{Review of Geophysics and Space Physics}	
\def\RS{Radio Science}					
\def\SP{Signal Processing}
\def\siamAM{SIAM Journal of Applied Mathematics}
\def\siamCO{SIAM Journal of Control and Optimization}
\def\siamJO{SIAM Journal of Optimization}		
\def\siamMA{SIAM Journal of Mathematical Analysis}
\def\siamNA{SIAM Journal of Numerical Analysis}
\def\siamO{SIAM Journal of Optimization}
\def\siamR{SIAM Review}
\def\SSR{Stochastics and Stochastics Reports}           
\def\TPA{Theory of Probability and its Applications}	
\def\TMK{Technometrics}
\def\TS{Traitement du Signal}
\def\UMB{Ultrasound in Medecine and Biology}
\def\US{Ultrasonics}
\def\USI{Ultrasonic Imaging}

%
\def\jan{janvier }
\def\feb{f\'evrier }
\def\mar{mars }
\def\apr{avril }
\def\may{mai }
\def\jun{juin }
\def\jul{juillet }
\def\aug{ao\^ut }
\def\sep{septembre }
\def\oct{octobre }
\def\nov{novembre }
\def\dec{d\'ecembre }
\def\Jan{January }	
\def\Feb{February }
\def\Mar{March }
\def\Apr{April }
\def\May{May }
\def\Jun{June }
\def\Jul{July }
\def\Aug{August }
\def\Sep{September }
\def\Oct{October }
\def\Nov{November }
\def\Dec{December }
\def\sub{soumis \`a }
\bibliographystyle{maxent}
\bibliography{gpibase,gpipubli,amd}
}

\end{document}